\documentclass[%
 aip,
 amsmath,amssymb,
 reprint,%
]{revtex4-1}

\usepackage{graphicx}
\usepackage{dcolumn}
\usepackage{bm}

\usepackage[utf8]{inputenc}
\usepackage[T1]{fontenc}
\usepackage{mathptmx}
\usepackage{color}
\usepackage[normalem]{ulem}

\begin{document}

\preprint{AIP/123-QED}

\title{One-Dimensional Edge Contact to Encapsulated MoS$_2$ with a Superconductor}

\author{A. Seredinski}
\thanks{These authors contributed equally to this work}
\affiliation{School of Sciences and Humanities, Wentworth Institute of Technology, Boston, MA 02115}
\affiliation{Department of Physics, Duke University, Durham, NC, 27708}

\author{E.G. Arnault}
\thanks{These authors contributed equally to this work}
\affiliation{Department of Physics, Duke University, Durham, NC, 27708}

\author{V.Z. Costa}
\affiliation{Department of Physics and Astronomy, San Francisco State University, San Francisco, CA 94132}

\author{L. Zhao}
\affiliation{Department of Physics, Duke University, Durham, NC, 27708}

\author{T.F.Q. Larson}
\affiliation{Department of Physics, Duke University, Durham, NC, 27708}

\author{K. Watanabe}
\affiliation{National Institute for Materials Science, Namiki 1-1, Tsukuba, Ibaraki 305-0044, Japan}

\author{T. Taniguchi}
\affiliation{National Institute for Materials Science, Namiki 1-1, Tsukuba, Ibaraki 305-0044, Japan}

\author{F. Amet}
\affiliation{Department of Physics and Astronomy, Appalachian State University, Boone, NC 28607}

\author{A.K.M. Newaz}
\affiliation{Department of Physics and Astronomy, San Francisco State University, San Francisco, CA 94132}

\author{G. Finkelstein}
\affiliation{Department of Physics, Duke University, Durham, NC, 27708}

\date{\today}

\begin{abstract}
Establishing ohmic contact to van der Waals semiconductors such as MoS$_2$ is crucial to unlocking their full potential in next-generation electronic devices. Encapsulation of few layer MoS$_2$ with hBN preserves the material’s electronic properties but makes electrical contacts more challenging. Progress toward high quality edge contact to encapsulated MoS$_2$ has been recently reported. Here, we evaluate a contact methodology using sputtered MoRe, a Type II superconductor with a relatively high critical field and temperature commonly used to induce superconductivity in graphene. We find that the contact transparency is poor and that the devices do not support a measurable supercurrent down to 3 Kelvin, which has ramifications for future fabrication recipes.
\end{abstract}

\maketitle
 Soon after the isolation of monolayer graphene, it was found that mono- and few-layer crystals could be isolated from transition metal dichalcogenides (TMDs) \cite{Novoselov2005}. TMDs host an array of interesting phenomena including superconductivity, charge density waves, and quantum spin Hall states \cite{Manzeli2017}. Among the library of TMDs, molybdenym disulfide (MoS$_2$) has attracted attention due to its layer-dependent band structure \cite{Mak2010,Lee2010}, high mobility \cite{Radisavljevic2011,Kim2012,Baugher2013}, large spin-orbit interaction \cite{Zhu2011,Xiao2012,Kosmider2013}, and gate-induced superconductivity \cite{Ye2012,Taniguchi2012,Lu2015,Costanzo2016}. Encapsulation of MoS$_2$ with hexagonal boron nitride (hBN) both protects it from atmosphere and separates it from sources of disorder \cite{Lee2015,Cao2015}. However, due to Schottky barriers, a readily formed oxide layer, and the fabrication challenges that come along with encapsulation, ohmic contact to hBN/MoS$_2$/hBN heterostructures has proven difficult. 
 
 \begin{figure}
    \centering
    \includegraphics[width=\linewidth]{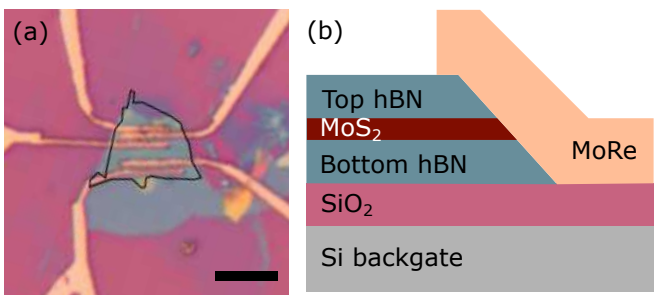}
    \caption{(a) Optical image of the first sample. The black outline shows the location of the encapsulated MoS$_2$. Scale bar 5 $\mu$m. (b) Schematic side view of the one-dimensional edge contact between the encapsulated MoS$_2$ and the sputtered MoRe (not to scale).}
    \label{fig:sample}
\end{figure}

Low temperature ohmic contact of normal metals to encapsulated MoS$_2$ has been achieved through workfunction engineering
\cite{Cui2017} as well as intervening graphene layers \cite{Lee2015,Cui2015}. Recently, progress has been made in one-dimensional edge contact to MoS$_2$ with normal metal through in situ Ar+ sputtering \cite{Jain2019,Cheng2019}. It would be highly desirable to develop superconducting edge contact to MoS$_2$, which could enable the study of the Josephson junction physics taking advantage of MoS$_2$'s spin-orbit and spin-valley couplings.

In this work we make one-dimensional edge contact to encapsulated MoS$_2$ using molybdenum-rhenium (MoRe), a Type II superconductor known to form high transparency contact to MoS$_2$ for a 2D interface \cite{Island2016}. We utilize a recipe known to make ohmic edge contacts to hBN-encapsulated graphene \cite{Calado2015,Borzenets2016}. Our measurements show low transparency contact to MoS$_2$ that is improved neither by Ar+ sputtering pre-treatment of the contact interfaces nor by annealing. These results indicate the probable presence of interfacial tunnel barriers. This result may prove informative for groups developing hybrid samples made of van der Waals heterostructures with superconducting contacts.

We study two MoS$_2$ devices encapsulated within hBN. Both samples are contacted by several MoRe electrodes, which define a series of Josephson junctions of different lengths. The first device uses bilayer MoS$_2$, while the second device uses monolayer MoS$_2$. Figure \ref{fig:sample} shows an optical image of the first device as well as a schematic view of the one-dimensional edge contact between the MoS$_2$ and MoRe, created via reactive ion etching and sputtering. The second device underwent an in situ Ar+ sputtering pre-treatment immediately before MoRe deposition. 

\begin{figure}
    \centering
    \includegraphics[width=\linewidth]{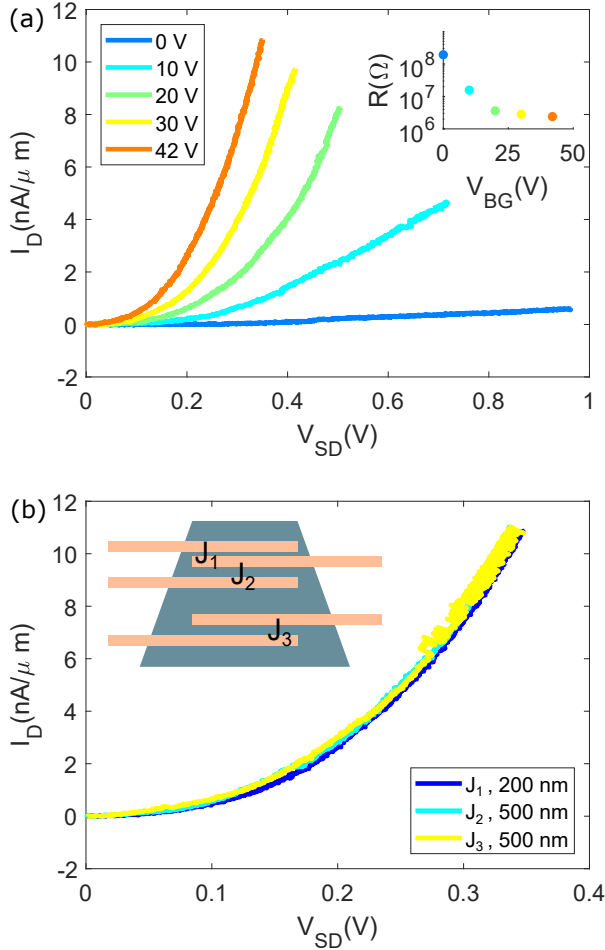}
    \caption{(a) Gate voltage dependence of the $I-V$ characteristics in a 200 nm long, 5 $\mu$m wide junction on the first device ($J_1$). Inset: resistance at high $V_{SD}$ for each gate voltage. The junction is seen to be highly resistive across applied gate and bias voltages, and no signs of superconducting behavior are visible. (b) $I-V$ curves for junctions $J_{1-3}$ of the first sample at $V_{BG}=42$ V. There is no significant difference between the 200 nm and 500 nm long junctions, indicating that the current is limited by the contacts. Inset: top-down schematic of the sample with $J_{1-3}$ labeled.}
    \label{fig:gate}
\end{figure}

Both van der Waals heterostructures were assembled from mechanically exfoliated flakes using a dry transfer technique utilizing a polyethylene terephthalate stamp. Polymer residue was removed by immersion in dichloromethane for one hour at 70 $^{\circ}$C followed by several hours at room temperature.

The one-dimensional interface between the MoS$_2$ and the MoRe was prepared via standard electron-beam lithography techniques, reactive ion etching (RIE), and sputtering. RIE consisted of three steps, all carried out with a process pressure of 10$^{-1}$ Torr. First, a ten second CHF$_3$ / O$_2$ (10:1 flow rate ratio) step removed leftover e-beam resist (PMMA) residue from the top surface of the heterostructure. This was followed by a ten second SF$_6$ process to etch through the top hBN. Finally, a ten second CF$_4$ step was used to etch the MoS$_2$ in the contact region. While a CF$_4$ etch is a typical process for MoS$_2$, SF$_6$ may itself be sufficient \cite{Jain2019}. In order to limit the device’s exposure to atmosphere, and so the formation of MoO$_x$ along the interface, the device was not removed from the system and imaged between these steps.

The devices had minimal exposure to air before being transferred to the sputtering system. The second sample was treated with Ar+ sputtering 
before metal deposition to refresh the contact interface. The chamber was pumped to a pressure of $\sim 10^{-8}$ Torr and 100 nm of MoRe (50-50\% by weight) was sputtered on both devices. To minimize processing, the Josephson junctions were not shaped with further etching, so the flakes of MoS$_2$ continue beyond the boundaries of the junctions. This is visible in Figure 1a, which shows an optical image of the first device.

The samples are cooled in a closed-cycle cryocooler with a base temperature of 3 K.  Unless otherwise noted, a voltage $V_{applied}$ is applied to the junction in series with a protective $R_S=$10 M$\Omega$ resistor. The drain current, $I_D$ is measured, and the source-drain voltage is calculated as $V_{SD}=V_{applied}-R_SI_D$; as a result the curves in Figures 2 and 3 have different horizontal extent.

Figure \ref{fig:gate}a shows the effects of electrostatic gating on the $I-V$ curves of a 200 nm long and 5 $\mu$m wide junction made on the first device. The gate voltage ($V_{BG}$) increases the Fermi level in the MoS$_2$, causing it to approach the conduction band. We observe that for increasing $V_{BG}$, the threshold of $V_{SD}$ required to achieve a linear slope decreases. Figure 2b demonstrates the $I-V$ curves measured for three junctions of different length at the maximal gate voltage of 42 V. (See the schematic in the inset: $J_1$ is 200 nm long, and $J_{2,3}$ are 500 nm long.) 
It is clear that 1) the curves show no significant length dependence, indicating that the current is limited by the contact barriers; and 2) the measurements are consistent between the three junctions, indicating uniform properties of the contacts. These initial measurements are consistent with the presence of barriers (likely Schottky barriers) at the interfaces \cite{Jain2019}. At the highest gate voltage (42 V) the resistance is 2.4 M$\Omega$, corresponding to the the contact resistance of $R_c\approx$ 6 M$\Omega$$\cdot$$\mu$m.

Due to this high contact resistance, we next anneal the sample at 200$^{\circ}$C for 17 hours in a vacuum of $10^{-6}$ mbar. Annealing processes have been shown to decrease contact resistance in similar devices. This may be due to a host of phenomena which change the bonding or structure at the interface \cite{Jain2019}. In this study, the annealing resulted in higher contact resistance, with an increase of as much as 40\% at high bias and $V_{BG}=42$ V. This decrease in contact quality may be due to the MoRe reflowing away from the contact edge, as seen in gold junctions without an additional metal sticking layer \cite{Jain2019}. 

\begin{figure}
\centering
\includegraphics[width=\linewidth]{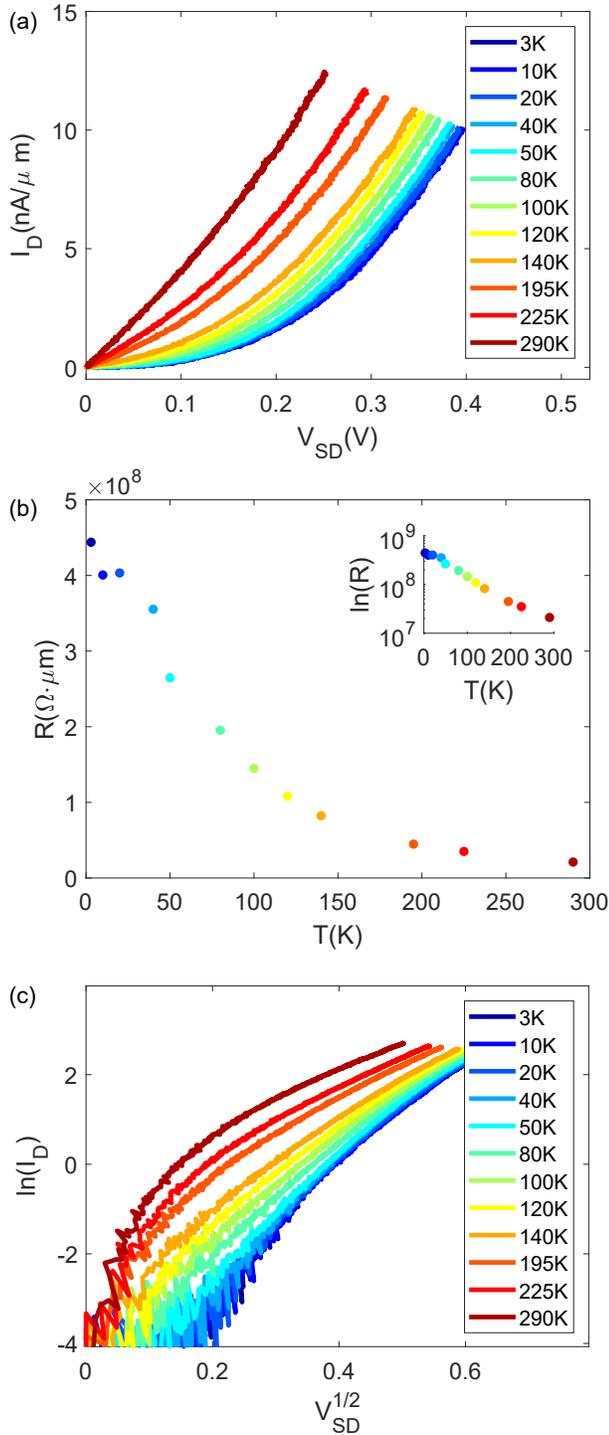}

    \caption{Temperature dependencies measured in the 200 nm long, 5 $\mu$m wide junction in the first device. (a) Post-anneal $I-V$ characteristics. (b) Low bias ($V_{SD}=0.05$ V) resistance $R$, plotted in linear and (inset) log scale, which shows $R$ decaying with temperature. $V_{BG}=42$ V throughout. (c) $\ln(I_D)$ vs $(V_{SD}/\mathrm{Volt})^{1/2}$ plot of the same data showing an approximately linear relationship in the intermediate temperature range. This is consistent with thermionic transport across the contact interfaces.}
    \label{fig:temp}
\end{figure}

We study the behavior of the junction as a function of temperature to gain insight into the poor contact quality. Figure \ref{fig:temp}a plots the $I-V$ characteristics of the same junction from 3 to 290 K. A clear reduction in low-bias resistance spanning more than a decade is seen as the temperature rises (Figure 3b). Such behavior is consistent with thermionic transport across a barrier. This interpretation is supported by an approximately linear relation between the log of the current and the square root of the bias voltage in the device (Figure 3c) as expected, e.g., for a triangular Schottky barrier~\cite{Sze2006}. This relation breaks down for low bias voltages at higher temperatures.

\begin{figure}
    \centering
    \includegraphics[width=\linewidth]{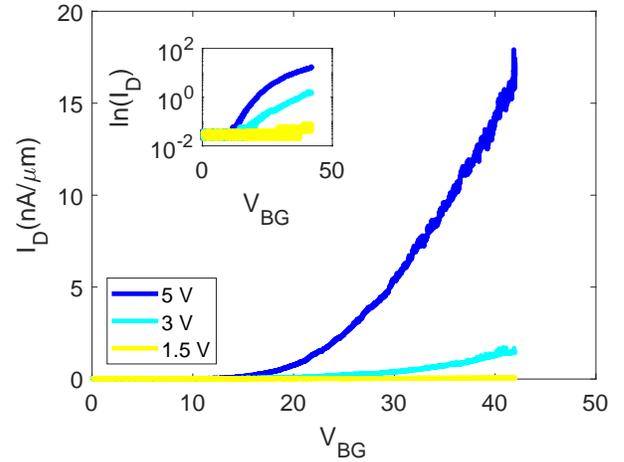}
    \caption{Current vs $V_{BG}$ sweeps measured in a 500 nm long by 5 $\mu$m wide junction in the second device following the annealing, which show the induced highly resistive behavior. The three curves correspond to $V_{SD}=$ 1.5, 3, and 5 V. Inset: the same data in log scale.}
    \label{fig:bias}
\end{figure}

Due to the contact characteristics of this device, we study a second device utilizing Ar+ sputtering immediately prior to the deposition of the MoRe contacts, focusing on a 500 nm long and 5 $\mu$m wide junction. Despite this change in deposition parameters and an overnight anneal at 300$^{\circ}$C in $10^{-6}$ mbar, this second device also displays high contact resistances at low temperature. Utilizing a direct voltage biasing scheme without a 10 M$\Omega$ series resistor, we measure gate sweeps for different $V_{SD}$ (Figure \ref{fig:bias}). Even at the highest applied V$_{SD}$=5 V, the currents supported by the junction are orders of magnitude lower than comparable or longer junctions made with both top contacts \cite{Liu2015,Smithe2016} and high quality normal metal edge contacts \cite{Jain2019,Cheng2019}.

In summary, we tested a methodology for making one-dimensional edge contact to encapsulated MoS$_2$ with MoRe, and found high contact resistances on the order of M$\Omega\cdot\mu$m. This contact was not improved by annealing at 200-300 $^{\circ}$C. In situ Ar+ sputtering of the interface before the deposition of MoRe also did not improve the contact quality. We conclude that the presence of tunnel barriers limits the performance of these devices. The lack of length dependence, consistency between different junctions,  insensitivity to Ar+ pre-cleaning, and the lack of improvement upon annealing all point to the presence of intrinsic Schottky barriers at the interfaces.

Higher transparency contacts may be achieved in the future by replacing MoRe with superconductors having a significantly higher or lower work function. Nevertheless, the current contact recipe could support the use of MoS$_2$ in more complex superconducting heterostructures. Namely, TMDs, including MoS$_2$\cite{Safeer2019}, are already used to induce the spin-orbit coupling in graphene \cite{Wang2016,Island2019}. One can extend these studies to Josephson junctions by making superconducting contacts that would selectively contact the graphene but not the TMD layer. In this context, our work establishes an order of magnitude estimate for the (very small) current expected to be shunted through an MoS$_2$ layer in such a complex van der Waals heterostructure.

\begin{acknowledgments}
A.S., E.G.A, T.F.L., L.Z., and G.F. acknowledge support by the Office of Basic Energy Sciences, U.S. Department of Energy, under Award de-sc0002765. V.Z.C. and A.K.M.N. acknowledge support from the National Science Foundation Grant ECCS-1708907 and Department of Defense Award (ID: 72495RTREP). K.W. and T.T.acknowledge the Elemental Strategy Initiative conducted by the MEXT, Japan and the CREST (JPMJCR15F3), JST. F.A. acknowledges the ARO under Award W911NF-16-1-0132. This work was performed in part at the Duke University Shared Materials Instrumentation Facility (SMIF), a member of the North Carolina Research Triangle Nanotechnology Network (RTNN), which is supported by the National Science Foundation (Grant ECCS-1542015) as part of the National Nanotechnology Coordinated Infrastructure (NNCI).
\end{acknowledgments}




\linespread{2}
\bibliography{MoSbib2}


\end{document}